\documentclass{sigchi-ext}
\pdfoutput=1
\usepackage[T1]{fontenc}
\usepackage{textcomp}
\usepackage[scaled=.92]{helvet} % for proper fonts
\usepackage{graphicx} % for EPS use the graphics package instead
\usepackage{balance}  % for useful for balancing the last columns
\usepackage{booktabs} % for pretty table rules
\usepackage{ccicons}  % for Creative Commons citation icons
\usepackage{ragged2e} % for tighter hyphenation

\usepackage{xcolor} % for colored texts

\usepackage{enumitem}
\usepackage[utf8]{inputenc}
\usepackage[acronym]{glossaries}
\glsdisablehyper
\usepackage{cuted}
\usepackage{capt-of}

\usepackage{cite}

\usepackage{xspace}

\newacronym{VR}{VR}{virtual reality}
\newacronym{MR}{MR}{mixed reality}
\newacronym{AR}{AR}{augmented reality}
\newacronym{UX}{UX}{user experience}
\newacronym{UI}{UI}{user interface}
\newacronym{GUI}{GUI}{graphical user interface}
\newacronym{DOF}{DoF}{Degrees of Freedom}
\newacronym{CAD}{CAD}{computer\hbox{-}aided design}
\newacronym{IMU}{IMU}{Inertial Measurement Unit}
\newacronym{WIM}{WIM}{World In Miniature}
\newacronym{SUS}{SUS}{System Usability Scale}
\newacronym{PDA}{PDA}{Personal Digital Assistant}
\newacronym{EMR}{EMR}{Electromagnetic Resonance}
\newacronym{SE}{SE}{social engineering}
\newacronym{PT}{PT}{penetration tester}
\newacronym{NPC}{NPC}{non-player character}
\newacronym[plural=CAVEs,firstplural=Cave Automatic Virtual Environments (CAVE)]{CAVE}{CAVE}{Cave Automatic Virtual Environment}
\newacronym{STEM}{STEM}{Science, Technology, Engineering and Mathematics}
\newacronym[plural=HMDs,firstplural=head\hbox{-}mounted displays (HMDs)]{HMD}{HMD}{head\hbox{-}mounted display}
\newacronym[plural=RCPs,firstplural=Rich Client Platforms (RCP)]{RCP}{RCP}{Rich Client Platform}
\newacronym[plural=IVs,firstplural=Independent Variables (IV)]{IV}{IV}{Independent Variable}
\newacronym[plural=DVs,firstplural=Dependent Variables (DV)]{DV}{DV}{Dependent Variable}

\def\plaintitle{Discussing the Risks of Adaptive Virtual Environments for User Autonomy} 
\def\emptyauthor{}
\def\plainkeywords{serious games; educational games; virtual reality; adaptivity; player state assessment; stealth assessment; user autonomy; privacy;}

\title{Discussing the Risks of Adaptive Virtual Environments for User Autonomy}

\numberofauthors{6}
\author{%
  \alignauthor{%
    \textbf{Tobias Drey}\\
    \affaddr{Institute of Media Informatics} \\
    \affaddr{Ulm University, Ulm, Germany} \\
    \email{tobias.drey@uni-ulm.de} }\alignauthor{%
    \textbf{Enrico Rukzio}\\
    \affaddr{Institute of Media Informatics} \\
    \affaddr{Ulm University, Ulm, Germany} \\
    \email{enrico.rukzio@uni-ulm.de} } \vfil \alignauthor{%
    } 
    }

\definecolor{linkColor}{RGB}{6,125,233}
\hypersetup{%
  pdftitle={\plaintitle},
  pdfauthor={\emptyauthor},
  pdfkeywords={\plainkeywords},
  bookmarksnumbered,
  pdfstartview={FitH},
  colorlinks,
  citecolor=black,
  filecolor=black,
  linkcolor=black,
  urlcolor=linkColor,
  breaklinks=true,
}

\copyrightinfo{\scriptsize Presented at the workshop “Exploring Potentially Abusive Ethical, Social and Political Implications of Mixed Reality Research in HCI”, CHI '20, April 25-30, Honolulu, HI, USA. \\
Copyright held by author(s).}

\begin{document}

\maketitle

% Uncomment to disable hyphenation (not recommended)
% https://twitter.com/anjirokhan/status/546046683331973120
%\RaggedRight{} 

% Do not change the page size or page settings.
\begin{abstract}
%Motivation
Adaptive virtual environments are an opportunity to support users and increase their flow, presence, immersion, and overall experience.
Possible fields of application are adaptive individual education, gameplay adjustment, professional work, and personalized content.
%Problem
But who benefits more from this adaptivity, the users who can enjoy a greater user experience or the companies or governments who are completely in control of the provided content.
While the user autonomy decreases for individuals, the power of institutions raises, and the risk exists that personal opinions are precisely controlled.
%Solution
In this position paper, we will argue that researchers should not only propose the benefits of their work but also critically discuss what are possible abusive use cases.
%Method
Therefore, we will examine two use cases in the fields of professional work and personalized content and show possible abusive use.
%Contribution

\end{abstract}

\keywords{\plainkeywords}

% ACM Classfication

\begin{CCSXML}
<ccs2012>
   <concept>
       <concept_id>10002978.10003029.10011150</concept_id>
       <concept_desc>Security and privacy~Privacy protections</concept_desc>
       <concept_significance>500</concept_significance>
       </concept>
   <concept>
       <concept_id>10003120.10003121.10003124.10010866</concept_id>
       <concept_desc>Human-centered computing~Virtual reality</concept_desc>
       <concept_significance>300</concept_significance>
       </concept>
   <concept>
       <concept_id>10010405.10010489.10010491</concept_id>
       <concept_desc>Applied computing~Interactive learning environments</concept_desc>
       <concept_significance>300</concept_significance>
       </concept>
 </ccs2012>
\end{CCSXML}

\ccsdesc[500]{Security and privacy~Privacy protections}
\ccsdesc[300]{Human-centered computing~Virtual reality}
\ccsdesc[300]{Applied computing~Interactive learning environments}

% Print the classficiation codes
\printccsdesc
%Please use the 2012 Classifiers and see this link to embed them in the text: \url{https://dl.acm.org/ccs/ccs_flat.cfm}

\section{Introduction}
%Motivation
Adaptivity provides an opportunity to provide individual~\gls{UX} and increase flow, presence, and immersion.
Related work has shown, that behavioral indicators and biometric data for \textit{hand} \cite{glowinski2011toward, piana2014real} and \textit{foot movements} \cite{pan2019foot, felberbaum2018better}, as well as \textit{gaze behavior} \cite{soler2017proposal, lankes2019lost,Sundstedt2013}, can be used to adapt virtual environments and \gls{VR}~\cite{LBW,10.1145/3306212.3328138}.
The use of \gls{VR} \glspl{HMD} provides the possibility to increase these effects by encapsulating users to their own environment.

%Problem/Solution
The intention behind adaptivity is that it is suitable to increase the effectiveness of education and professional work, the overall gameplay experience of video games, and to introduce perfectly matching personal content like a news feed.
But the risk exists that adaptivity and all its benefits come at the cost of losing freedom and control.
Whenever a machine makes a decision and adapts a system for us to our possible favor, then it also withholds information or possible solutions from us.
On the positive side, this can be beneficial as it helps us to filter a lot of unnecessary information so that we have more time to concentrate on important things.
On the negative side, it provides the possibility to hide information and install unnoticed censorship.
This misuse is often ignored when new adaptive systems are developed and evaluated as researchers are excited about what they did, but a profound discussion in a paper should also consider negative and possible abusive aspects.

%Method (Implemented/Explored)
%planned Contribution (only introduction)
In this position paper, we discuss two use cases and show how good and helpful approaches can be abused.
We will consider a professional work \gls{VR} application and adaptive personalized content based on user profiles.

\section{Discussion of Abusive User Monitoring}
\subsection{Problem Definition}
Adaptive educational games help to provide individual learning experiences by constantly monitoring the user~\cite{doi:10.1111/jcal.12416}. 
This is beneficial for the learning context.
Techniques were developed to assess different capabilities of the user like spatial abilities~\cite{Shepard701} or problem understanding~\cite{Clarke_Peel_Arnab_Morini_Keegan_Wood_2017}.
Most of these techniques have in common that they are completely unrecognized by the user.
Therefore, the risk exists that a user is monitored and assessed without knowledge.
\subsection{Adaptive Individual Work Environment}
One possible example is an employee working with \gls{CAD} \gls{VR} software.
Similar, as proposed by Drey et al.~\cite{LBW} and Bye et al.~\cite{10.1145/3306212.3328138}, this software monitors all body movements, including hands, feet, head, and even the gaze behavior.
Furthermore, the workflow is monitored too, and the personal capabilities were assessed based on measurements like spatial abilities.
This could then be automatically reported to the management which results in a possible promotion or a job downgrade.
Besides reporting to the management, the software could select the tasks for the employee based on the assessed skills.
This could be beneficial for the company as employees always work on tasks perfectly fitting to their skills, but it is also problematic as they have no possibility to improve their skills with challenging tasks.
When this adaptive preselection is done completely non-transparent, then this is a sort of censorship.
\subsection{Adaptive Personalized Content Based on User Profiles}
All major IT companies like Facebook, Google, Apple, or Microsoft create user profiles based on user activity~\cite{Areyoure32:online}.
These profiles can be enriched with data gained through adaptive games with leisure or educational content.
The methods proposed by Drey et al.~\cite{LBW} and Bye et al.~\cite{10.1145/3306212.3328138} provide new possibilities to create a much more complex user profile than today.
They tell a lot about how users understand and interact with their environment.
This data is used, at the moment, mostly for advertisement and content predictions.
Personally perceived interesting content is a positive thing, but it creates on the other side an individual bubble, which leads to a sort of automatic censorship.

The more data is collected, it gets more likely that there is only personalized content and no "uncensored" general one.
In a dystopic future, companies and governments use all data they are gathering from their users or citizens to monitor and assess them.
The techniques now developed for positive adaptive environments are used then to create censored environments.
Users cannot opt\hbox{-}out from this and do not know how they are monitored due to the use of unobtrusive techniques.
The public opinion is controlled as everyone gets only censored and personally fitted information.

The reason why no general and uncontrolled way to get information exists anymore is that newspapers and other independent organizations got bankrupt because users preferred their personalized information.
As censorship increased slowly during the time, they were not aware of how they were manipulated.
This was a slow, unobtrusive, but constant process until the power of the companies and governments was high enough to control public opinion.
\section{Conclusion}
The described two scenarios in the fields of a professional working environment and personalized content are clearly dystopic worst\hbox{-}case scenarios.
There are laws and regulations who should protect employees and citizens from permanent monitoring.
Nevertheless, it is important that everyone is aware of what is technically possible to control whether laws and regulations are kept.
Therefore, we argue that researchers should consider possible abusive use cases as well and point out this to the readers of their papers.
Whenever this is not done sufficiently by the authors themselves, this should be done by the community by raising concerns.

At the workshop, I would like to talk about these dystopic scenarios and discuss who of the participants share my concerns about adaptive virtual environments.
Furthermore, I would like to discuss possible actions to prevent abusive use.

\section{Acknowledgments}
This work was conducted within the project AuCity 2, funded by the Federal Ministry of Education and Research (BMBF).

\balance{} 

\bibliographystyle{SIGCHI-Reference-Format}
\bibliography{extended-abstract}

%%% -*-BibTeX-*-
%%% Do NOT edit. File created by BibTeX with style
%%% ACM-Reference-Format-Journals [18-Jan-2012].

\begin{thebibliography}{00}

%%% ====================================================================
%%% NOTE TO THE USER: you can override these defaults by providing
%%% customized versions of any of these macros before the \bibliography
%%% command.  Each of them MUST provide its own final punctuation,
%%% except for \shownote{}, \showDOI{}, and \showURL{}.  The latter two
%%% do not use final punctuation, in order to avoid confusing it with
%%% the Web address.
%%%
%%% To suppress output of a particular field, define its macro to expand
%%% to an empty string, or better, \unskip, like this:
%%%
%%% \newcommand{\showDOI}[1]{\unskip}   % LaTeX syntax
%%%
%%% \def \showDOI #1{\unskip}           % plain TeX syntax
%%%
%%% ====================================================================

\ifx \showCODEN    \undefined \def \showCODEN     #1{\unskip}     \fi
\ifx \showDOI      \undefined \def \showDOI       #1{{\tt DOI:}\penalty0{#1}\ }
  \fi
\ifx \showISBNx    \undefined \def \showISBNx     #1{\unskip}     \fi
\ifx \showISBNxiii \undefined \def \showISBNxiii  #1{\unskip}     \fi
\ifx \showISSN     \undefined \def \showISSN      #1{\unskip}     \fi
\ifx \showLCCN     \undefined \def \showLCCN      #1{\unskip}     \fi
\ifx \shownote     \undefined \def \shownote      #1{#1}          \fi
\ifx \showarticletitle \undefined \def \showarticletitle #1{#1}   \fi
\ifx \showURL      \undefined \def \showURL       #1{#1}          \fi

\bibitem{Areyoure32:online}
 2020.
\newblock Are you ready? This is all the data Facebook and Google have on you |
  Dylan Curran | Opinion | The Guardian.
\newblock   (Feb 2020).
\newblock
\showURL{%
\url{https://www.theguardian.com/commentisfree/2018/mar/28/all-the-data-facebook-google-has-on-you-privacy}}
\newblock
\shownote{(Accessed on 02/20/2020).}


\bibitem{10.1145/3306212.3328138}
{Kent Bye}, {Diane Hosfelt}, {Sam Chase}, {Matt Miesnieks}, {and} {Taylor
  Beck}. 2019.
\newblock \showarticletitle{The Ethical and Privacy Implications of Mixed
  Reality}. In {\em ACM SIGGRAPH 2019 Panels} {\em (SIGGRAPH ’19)}.
  Association for Computing Machinery, New York, NY, USA, Article Article 4, 2
  pages.
\newblock
\showISBNx{9781450363129}
\showDOI{%
\url{http://dx.doi.org/10.1145/3306212.3328138}}


\bibitem{Clarke_Peel_Arnab_Morini_Keegan_Wood_2017}
{Samantha~Jane Clarke}, {Daryl~J. Peel}, {Sylvester Arnab}, {Luca Morini},
  {Helen Keegan}, {and} {Oliver Wood}. 2017.
\newblock \showarticletitle{EscapED: A Framework for Creating Educational
  Escape Rooms and Interactive Games to For Higher/Further Education.}
\newblock {\em International Journal of Serious Games\/} {4}, 3 (Sep. 2017).
\newblock
\showDOI{%
\url{http://dx.doi.org/10.17083/ijsg.v4i3.180}}


\bibitem{LBW}
{Tobias Drey}, {Pascal Jansen}, {Fabian Fischbach}, {Julian Frommel}, {and}
  {Enrico Rukzio}. 2020.
\newblock \showarticletitle{Towards Progress Assessment for Adaptive Hints in
  Educational Virtual Reality Games}. In {\em Extended Abstracts of the 2020
  CHI Conference on Human Factors in Computing Systems} {\em (CHI ’20)}.
  Association for Computing Machinery, New York, NY, USA, Article Paper LBW324,
  9 pages.
\newblock
\showISBNx{978-1-4503-6819-3/20/04.}
\showDOI{%
\url{http://dx.doi.org/10.1145/3334480.3382789}}


\bibitem{felberbaum2018better}
{Yasmin Felberbaum} {and} {Joel Lanir}. 2018.
\newblock \showarticletitle{Better Understanding of Foot Gestures: An
  Elicitation Study}. In {\em Proceedings of the 2018 CHI Conference on Human
  Factors in Computing Systems}. ACM, 334.
\newblock


\bibitem{glowinski2011toward}
{Donald Glowinski}, {Nele Dael}, {Antonio Camurri}, {Gualtiero Volpe},
  {Marcello Mortillaro}, {and} {Klaus Scherer}. 2011.
\newblock \showarticletitle{Toward a minimal representation of affective
  gestures}.
\newblock {\em IEEE Transactions on Affective Computing\/} {2}, 2 (2011),
  106--118.
\newblock


\bibitem{lankes2019lost}
{Michael Lankes} {and} {Andreas Haslinger}. 2019.
\newblock \showarticletitle{Lost \& Found: Gaze-based Player Guidance Feedback
  in Exploration Games}. In {\em Extended Abstracts of the Annual Symposium on
  Computer-Human Interaction in Play Companion Extended Abstracts}. ACM,
  483--492.
\newblock


\bibitem{pan2019foot}
{Ye Pan} {and} {Anthony Steed}. 2019.
\newblock \showarticletitle{How Foot Tracking Matters: The Impact of an
  Animated Self-Avatar on Interaction, Embodiment and Presence in Shared
  Virtual Environments}.
\newblock {\em Frontiers in Robotics and AI\/}  {6} (2019), 104.
\newblock


\bibitem{piana2014real}
{Stefano Piana}, {Alessandra Stagliano}, {Francesca Odone}, {Alessandro Verri},
  {and} {Antonio Camurri}. 2014.
\newblock \showarticletitle{Real-time automatic emotion recognition from body
  gestures}.
\newblock {\em arXiv preprint arXiv:1402.5047\/} (2014).
\newblock


\bibitem{Shepard701}
{Roger~N. Shepard} {and} {Jacqueline Metzler}. 1971.
\newblock \showarticletitle{Mental Rotation of Three-Dimensional Objects}.
\newblock {\em Science\/} {171}, 3972 (1971), 701--703.
\newblock
\showISSN{0036-8075}
\showDOI{%
\url{http://dx.doi.org/10.1126/science.171.3972.701}}


\bibitem{soler2017proposal}
{Jose~L Soler-Dominguez}, {Jorge~D Camba}, {Manuel Contero}, {and} {Mariano
  Alca{\~n}iz}. 2017.
\newblock \showarticletitle{A proposal for the selection of eye-tracking
  metrics for the implementation of adaptive gameplay in virtual reality based
  games}. In {\em International Conference on Virtual, Augmented and Mixed
  Reality}. Springer, 369--380.
\newblock


\bibitem{Sundstedt2013}
{Veronica Sundstedt}, {Matthias Bernhard}, {Efstathios Stavrakis}, {Erik
  Reinhard}, {and} {Michael Wimmer}. 2013.
\newblock {\em Visual Attention and Gaze Behavior in Games: An Object-Based
  Approach}.
\newblock Springer London, London, 543--583.
\newblock
\showISBNx{978-1-4471-4769-5}
\showDOI{%
\url{http://dx.doi.org/10.1007/978-1-4471-4769-5_25}}


\bibitem{doi:10.1111/jcal.12416}
{Stefanie Vanbecelaere}, {Katrien Van~den Berghe}, {Frederik Cornillie},
  {Delphine Sasanguie}, {Bert Reynvoet}, {and} {Fien Depaepe}. 2019.
\newblock \showarticletitle{The effectiveness of adaptive versus non-adaptive
  learning with digital educational games}.
\newblock {\em Journal of Computer Assisted Learning\/} (2019).
\newblock
\showDOI{%
\url{http://dx.doi.org/10.1111/jcal.12416}}


\end{thebibliography}

\end{document}